\documentclass[aps,twocolumn,prl,showpacs]{revtex4}
\usepackage{epsfig}
\usepackage{graphicx}
\usepackage{amsfonts}
\usepackage[figuresright]{rotating} 
\usepackage{amssymb}
\usepackage{amsmath}

\def\avg#1{\langle#1\rangle}

\def\be{\begin{equation}} \def\ee{\end{equation}}
\def\bea{\begin{eqnarray}} \def\eea{\end{eqnarray}}

\def\pp{\parallel}

\begin{document}

\title{Orbital analogue of quantum anomalous Hall effect in $p$-band
systems}
\author{Congjun Wu}
\affiliation{Department of Physics, University of California, San Diego,
CA 92093}

\begin{abstract}
We investigate the topological insulating states of the $p$-band
systems in optical lattices induced by the onsite orbital angular
momentum polarization, which exhibit gapless edge modes in 
the absence of Landau levels.
This effect arises from the energy level splitting between the 
onsite $p_x+ip_y$ and $p_x-ip_y$ orbitals by rotating each optical 
lattice site around its own center.
At large rotation angular velocities, this model naturally reduces
to two copies of Haldane's quantum Hall model without Landau levels.
The distribution of Berry curvature in the momentum space and the 
quantized Chern numbers are calculated.
The experimental realization is also discussed.
\end{abstract}
\maketitle

The integer quantum Hall (QH) effect has generate tremendous research 
interests for several decades.
The precise quantization of the Hall conductance is due to the 
topologically non-trivial band structure characterized by the 
Thouless-Kohmoto-Nightingale-den Nijs (TKNN) number, or the Chern 
number \cite{thouless1982, kohmoto1985}.
The origin of the QH effect has also deep connections to the parity anomaly of
2D Dirac fermions \cite{jackiw1984, fradkin1986, haldane1988}.
Although breaking time-reversal (TR) symmetry is required, Landau levels 
(LL) are not necessary for the QH effect.
For example, Haldane \cite{haldane1988} constructed a QH model with 
average zero flux per unit cell but with complex-valued  hopping 
integrals. %which has non-zero Chern number and exhibits gapless edge states.
Recently, QH insulators have been generalized to the topological quantum 
spin Hall (QSH) insulators which keep time-reversal (TR) symmetry 
and are characterized by a Z$_2$ topological number \cite{
bernevig2006a, qi2008a, kane2005, hirsch1989, sheng2006, 
moore2007, roy2006}.
%Interaction effects in QSH insulators are investigated in
%\cite{wu2006, xu2006, lee2008, raghu2008}.
Excitingly, experimental evidence for the QSH insulating
states has been found \cite{konig2007, hsieh2008}.

Anomalous Hall (AH) effect describes the dependence of the Hall
current on the spin magnetization not the external magnetic field, 
whose mechanism has been debated for a long time,
including the anomalous velocity from the interband matrix element
\cite{karplus1954}, screw scattering\cite{smit1958}, and side jump
\cite{berger1970}.
Recently, a new perspective on the AH effect has been developed from the 
topological Berry curvature of the band structure which is a combined 
effect from spin-orbit coupling and spin polarization \cite {jungwirth2002, 
nagaosa2006}.
%This mechanism  describes the AH effect well in several materials 
%such as SrRuO$_3$, (In,Mn)As and (Ga,Mn)As systems.
Its quantum version, topological insulators arising from spin
magnetization has been proposed and investigated in semiconductor 
systems \cite{qi2006a, liu2008, onoda2003a}.

The current development of cold atom physics has provided another 
new opportunity to investigate the QH effect.
Several methods have been proposed including globally rotating the 
trap or optical lattice, or introducing effective gauge potential generated by 
laser beams \cite{ho2000, scarola2007, shao2008, umucalilar2008, zhu2006}.
However, the rotating angular velocity cannot be large enough otherwise 
the centrifugal potential will throw atoms away.
It is also difficult to make the light induced gauge potential strong
in a large region.
%Recently, a proposal to simulate Haldane's model
%has been suggested \cite{shao2008} by superpose a periodic light
%-induced gauge potential with a honeycomb optical lattice.
%However, it needs a high precision of alignment which might be difficult.

In this article, we propose an orbital analogue of the quantum anomalous
Hall (QAH) effect in solid state systems, i.e., the QAH effect
arising from orbital angular momentum polarization without LLs.
This can be achieved by rotating each optical site around its own center
which is an experimentally feasible technique \cite{gemelke2007}.
The lift of the degeneracy between $p_x\pm i p_y$ orbitals
gives rise to non-trivial topological band structures,
and provides a natural way to realize Haldane's model.
Increasing rotation angular velocity induces topological phase
transition by changing the Chern number of the band structure.
We also consider the QH effect arising with LLs in
such systems.

% an overall rotation in the
%$p$-orbital system in the honeycomb lattices.
%which has a combined effect of orbitals and Landau levels.

The experiment to rotate each site in the lattice around its own site 
center has been performed by Gemelke {\it et al.} \cite{gemelke2007}.
Electro-optic phase modulators are applied to the laser beams forming
the lattice, which results in a periodical overall translation of 
the lattice at a radio-frequency $\omega_{RF}$ but without the internal 
lattice distortions.
$\omega_{RF}$ is hundreds of times larger than the harmonic frequency
$\omega_L$ of each site, thus atoms only feel an averaged potential with a 
small distortion along the oscillation axis.
This axis can be controlled to rotate at an audio frequency $\Omega_z$,
which induces the rotation of each site around its own center at 
such a frequency.
$\Omega_z$ can be much larger than the overall parabolic trapping 
frequency and reach a few kilo-Hertz.
This technique has been applied in the triangular lattice described in 
Ref. \cite{gemelke2007}.

Let us consider to apply this  technique to the honeycomb lattice
which has been constructed quite some ago \cite{grynberg1993}.
We study the $p_{x,y}$-orbital band filled with spinless fermions
described in Ref. \cite{wu2007,wu2007a,wu2008} with the new
ingredient of rotation as
\bea
\label{eq:ham0}
H_0&=&t_{\pp} \sum_{ \vec{r} \in A} \big\{ p^\dagger_{\vec{r},i}
p_{\vec{r}+a\hat{e}_i,i}+h.c. \big\} -\mu\sum_{\vec{r}} n_{\vec{r}}, 
 \nonumber \\
H_L&=& i\Omega_z \sum_{\vec{r}} \big\{ p^\dagger_{\vec{r},x} p_{\vec{r},y}
- p^\dagger_{\vec{r},y} p_{\vec{r},x} \big\}, 
\eea
where $\hat{e}_{1,2}=\pm\frac{\sqrt{3}}{2}\hat{e}_x+\frac{1}{2}\hat{e}_y$
and $\hat{e}_3=-\hat{e}_y$ are the unit vectors pointing from a
site in the $A$-sublattice to its three neighbors in the $B$-sublattice;
$p_i\equiv (p_x\hat{e}_x+p_y\hat{e}_y)\cdot \hat{e}_i~(i=1\sim 3)$
are the projections of the $p$-orbitals along the $\hat e_i$ direction;
$\mu$ is the chemical potential;
$a$ is the nearest neighbor bond length.
Since there is no the overall lattice rotation, the vector potential
due to the Coriolis force and the centrifugal potential across the
entire lattices do not appear.
The effect is to break the degeneracy between $p_x\pm i p_y$
as described by $H_L$.
$\Omega_z$ can easily reach the order of the recoil energy $E_R$,
and $t_\pp$ can be tuned one order smaller than E$_R$ \cite{wu2007a},
thus we have a large flexibility of tuning $\Omega_z/t_\pp$.

\begin{figure}
\centering\epsfig{file=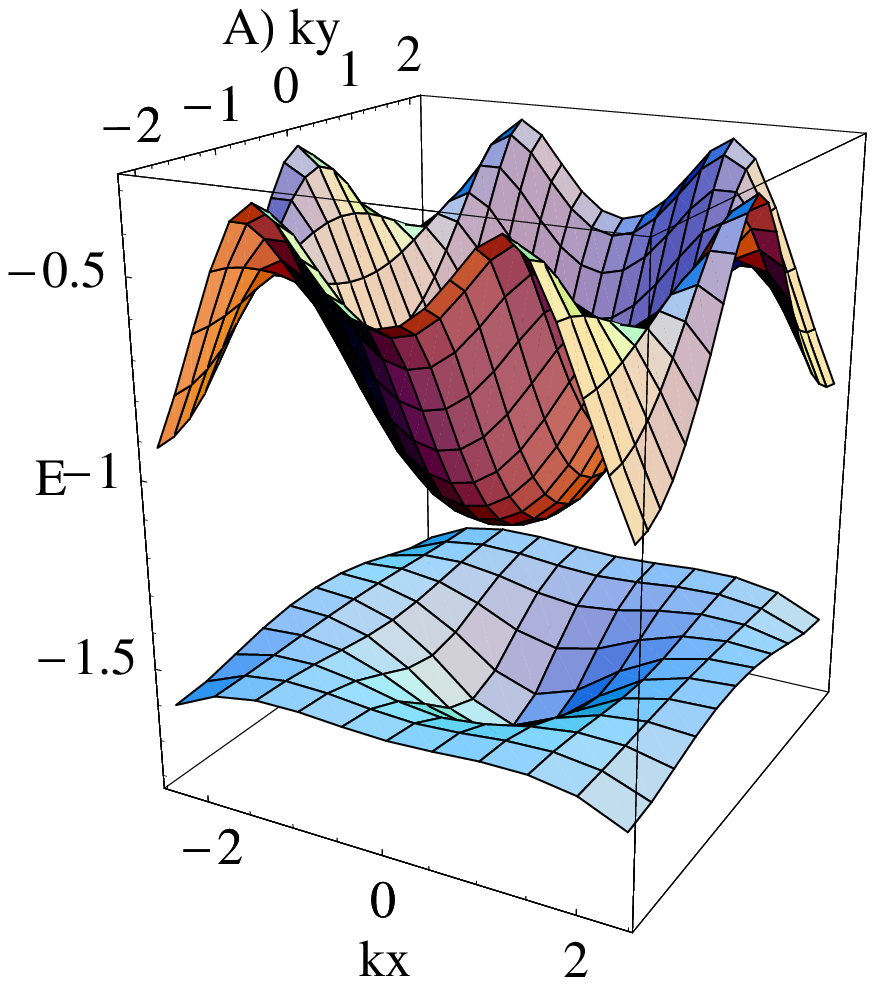,clip=1,width=0.45\linewidth,angle=0}
\centering\epsfig{file=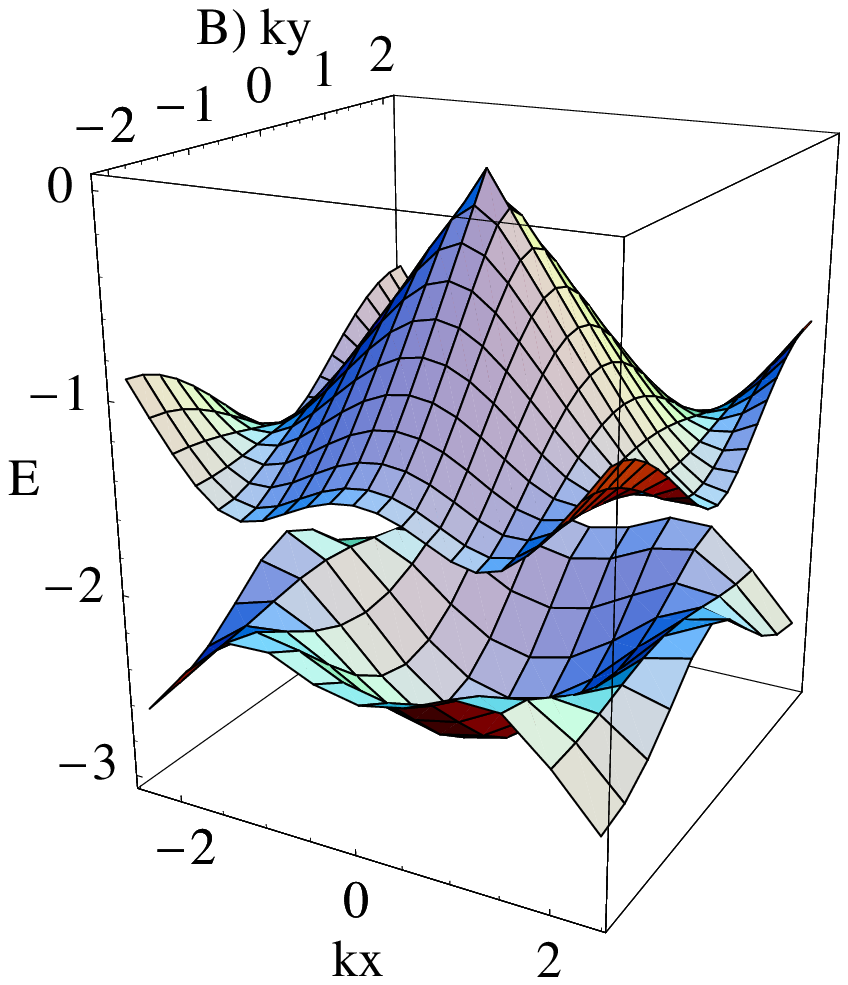,clip=1,width=0.45\linewidth,angle=0}
\centering\epsfig{file=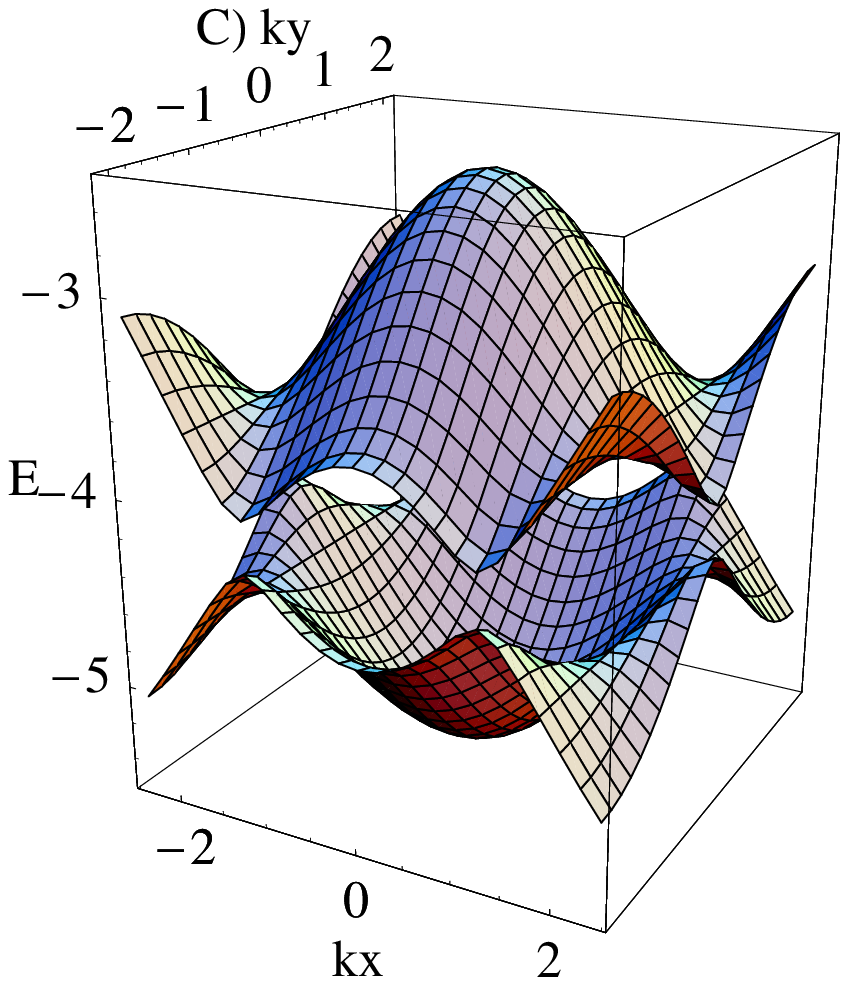,clip=1,width=0.45\linewidth,angle=0}
\centering\epsfig{file=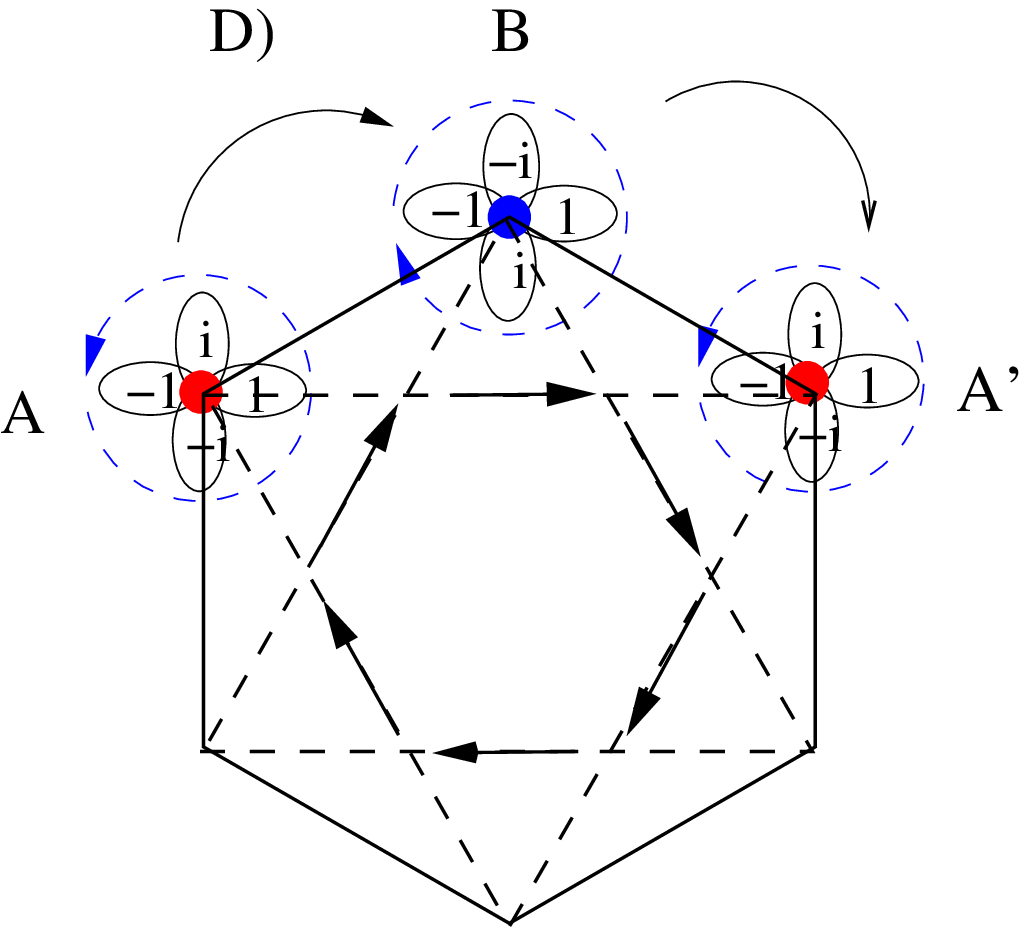,clip=1,width=0.52\linewidth,angle=0}
\caption{The band structure of Eq. \ref{eq:ham0} at $\Omega_z>0$ as 
shown in A, B and C. Only the lower two bands are presented, 
and the upper two are symmetric respect to the zero energy.
A) $\Omega_z/t_\pp=0.3$; B) $\Omega_z/t_\pp=1.5$ where a single 
gapless Dirac cone appears; C) $\Omega_z/t_\pp=3$ 
where two massive Dirac cones appear at $K_{1,2}$ between the lower 
two bands and also between the upper two bands;
D) The pattern of the induced next nearest neighbor hopping (complex-valued)
at $\Omega_z\gg t_\pp$, which is generated by the virtual hopping 
between orbitals with opposite chirality.
}
\label{fig:spectra}
\end{figure}

The band structure of Eq. \ref{eq:ham0} is presented as follows.
Under the chiral transformation $P$, i.e., $p_{r_A, x,y} \rightarrow 
p_{r_A, x,y}$, $p_{r_B, x,y}\rightarrow -p_{r_B, x,y}$, combined by 
the time-reversal transformation $T$, Eq. \ref{eq:ham0}
transforms as $ (TP)^{-1} (H_0+H_L) (TP)= -(H_0+H_L)$, thus its spectra are 
symmetric respect to the zero energy.
At $\Omega_z=0$ it exhibits two dispersive bands touching at
Dirac cones located at $K_{1,2}=(\pm \frac{4\pi}{3\sqrt 3 a}, 0)$
and other two flat bands \cite{wu2007,wu2007a}.
The dispersive bands touch the flat bands at the 
Brillouin zone (BZ) center $K_0=(0,0)$.
We define the 4-component spinor as
$\psi(\vec k)=(p_{Ax}(\vec k),p_{Ay}(\vec k), 
p_{Bx}(\vec k),p_{By}(\vec k))^T$, and
the two bases for the dispersive bands as
$\phi_1(\vec k)=\sqrt{\frac{2}{N_0}}\Big\{f_{12}(\vec k),\ \ \ 
\frac{1}{\sqrt 3}
(f_{23}(\vec k)-f_{31}(\vec k)),~ 0,~ 0\Big\}$, and
$\phi_2(\vec k)=\sqrt{\frac{2}{N_0}}
\Big\{0,~ 0,~ f^*_{12}(\vec k),\ \ \ \frac{1}{\sqrt 3}
(f^*_{23}(\vec k)-f^*_{31}(\vec k))\Big\},
$
where $f_{ij}=e^{i \vec k \cdot \hat e_i}
-e^{i \vec k \cdot \hat e_j}$ and $N_0(\vec k)$ is the
normalization factor.
At nonzero $\Omega_z$, gaps open between different bands as
depicted in Fig. \ref{fig:spectra} A, B, and C.
If $\Omega_z$ is small, the effective Hamiltonian close to
the Dirac points of $K_{1,2}$ can be written in the bases of
$\phi_1, \phi_2$ as
\bea
H_1(\vec k)=\left( \begin{array}{cc}
\frac{8\Omega_z}{\sqrt 3 N_0(\vec k)} \sum_i \sin \vec k \cdot \vec b_i
& -\frac{t_\pp}{2} \sum_i e^{-i \vec k \cdot \hat e_i}\\
-\frac{t_\pp}{2} \sum_i e^{i \vec k \cdot \hat e_i} &
\frac{-8\Omega_z}{\sqrt 3 N_0(\vec k)} \sum_i \sin \vec k \cdot \vec b_i
\end{array} \nonumber 
\right ), 
\label{eq:dirac1}
\eea
where $\vec b_i=\frac{1}{2}\epsilon_{ijk} (\hat e_j -\hat e_k)$ are the 
vectors connecting the next nearest neighbors.
The Dirac cones become gapped with $\Delta= \Omega_z$
and the masses are of the opposite sign at $K_{1,2}$.
%This is similar to Haldane's construction of the topological 
%insulator in the hexagonal lattice with the complex-valued next 
%nearest neighbor hopping term.
%In comparison, only the nearest neighbor hopping is involved 
%in Eq. \ref{eq:ham0}.
The bottom band is no longer flat at nonzero $\Omega_z$.
Its minimum at $K_0$ is pushed down by a value of $\frac{3}{2}
\Omega_z$ and that of the second band is pushed up by 
$\frac{3}{2} \Omega_z$.
This opens a gap of $3\Omega_z$.
A similar analysis applies to the top and the third bands.

As $\Omega_z$ approaches $\frac{3}{2}t_\pp$, the middle two bands at $K_0$
are pushed to zero from both up and below respectively, and form a single 
gapless Dirac cone in the BZ as depicted in Fig. \ref{fig:spectra} B.
Let us define another two  bases as 
$\phi_1^\prime=\frac{1}{2}\{1, i, -1, -i\}$, 
and $\phi_2^\prime=\frac{1}{2}\{1,-i, 1, -i\}$ for the Dirac cone
at $K_0$, and the Hamiltonian matrix becomes
\bea
H_2(\vec k)=\left( \begin{array}{cc}
-(\Omega_z-\frac{3}{2}t_\pp) & -\frac{3}{2} t_\pp (k_x+i k_y) \\
-\frac{3}{2} t_\pp (k_x-i k_y)& \Omega_z-\frac{3}{2}t_\pp
\end{array}
\right ).
\label{eq:dirac2}
\eea
We notice that a single Dirac cone of the chiral fermion is allowed in 
the 2D bulk lattice systems, which
actually does not contradict to the fermion doubling theory 
proved for 3D lattices \cite{nielsen1981}.

As $\Omega_z$ goes even larger, the lower and upper two pairs of
bands are projected into the single orbital bands of $p_x \pm i p_y$ 
respectively.
The lower two are described by the $p_x + i p_y$ orbital
with a nearest neighbor hopping of $\frac{t_\pp}{2}$.
Furthermore, a Haldane type next nearest neighbor hopping is generated 
as depicted in Fig. \ref{fig:spectra} D: 
one particle at site $A$ in the $p_x + i p_y$ orbital  hops to the 
high energy orbital of $p_x -i p_y$ at its nearest neighbor $B$, 
and hops back into the $p_x +i p_y$ state at the next nearest 
neighbor site $A^\prime$.
Along the directions indicted by arrows, this
hopping amplitude can be calculated from the second order
perturbation theory as $t_{nn}= t^2_\pp/(2\Omega_z) e^{i\frac{2}{3} \pi}.$  
As pointed out in Ref. \cite{haldane1988}, this generates two
massive Dirac cones with gap $\Delta=\frac{9}{2} t_{nn}$ 
at $\vec K_{1,2}$ of masses with opposite signs.

\begin{figure}
\centering\epsfig{file=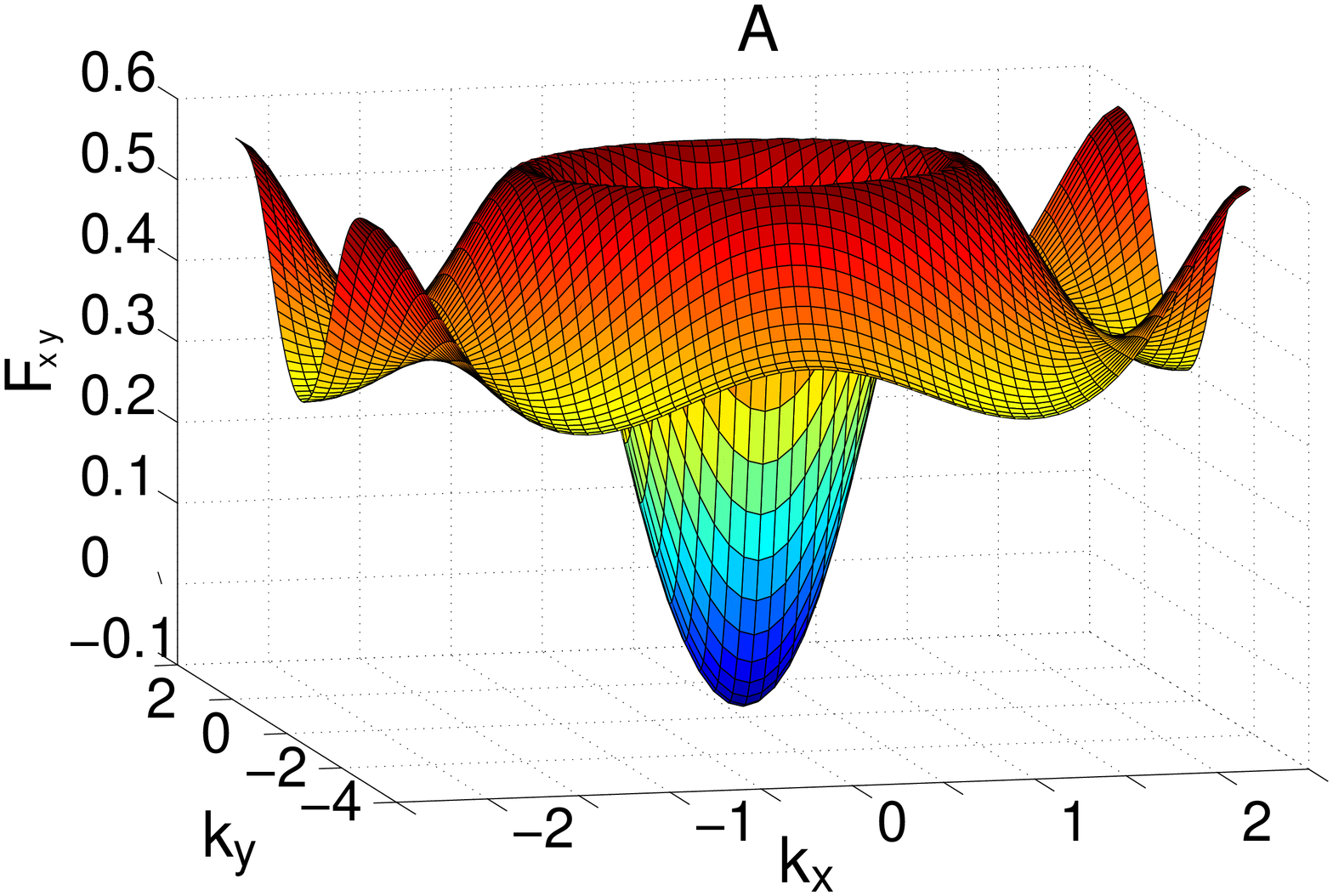,clip=1,width=0.49\linewidth,angle=0}
\centering\epsfig{file=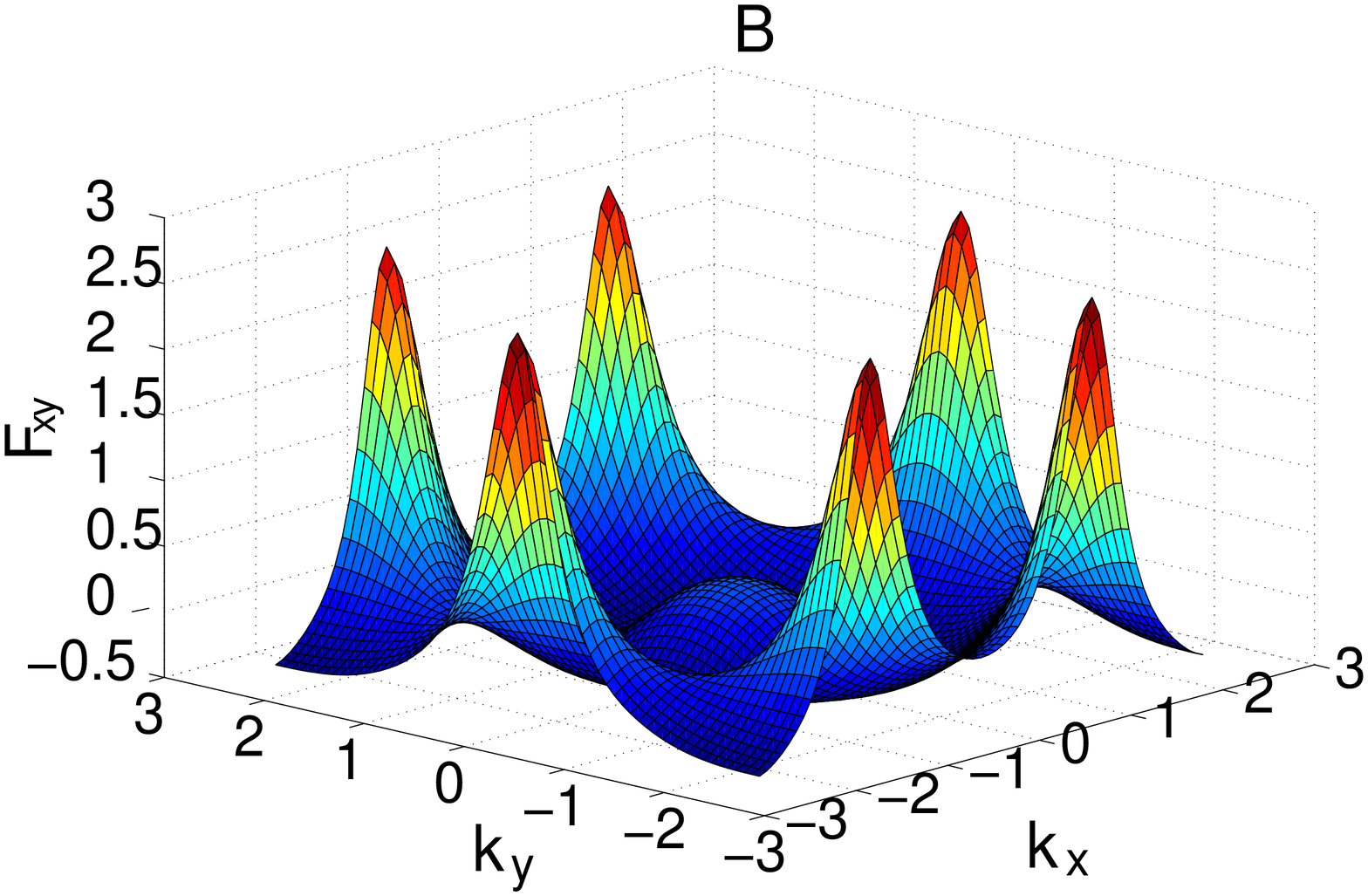,clip=1,width=0.49\linewidth,angle=0}
\centering\epsfig{file=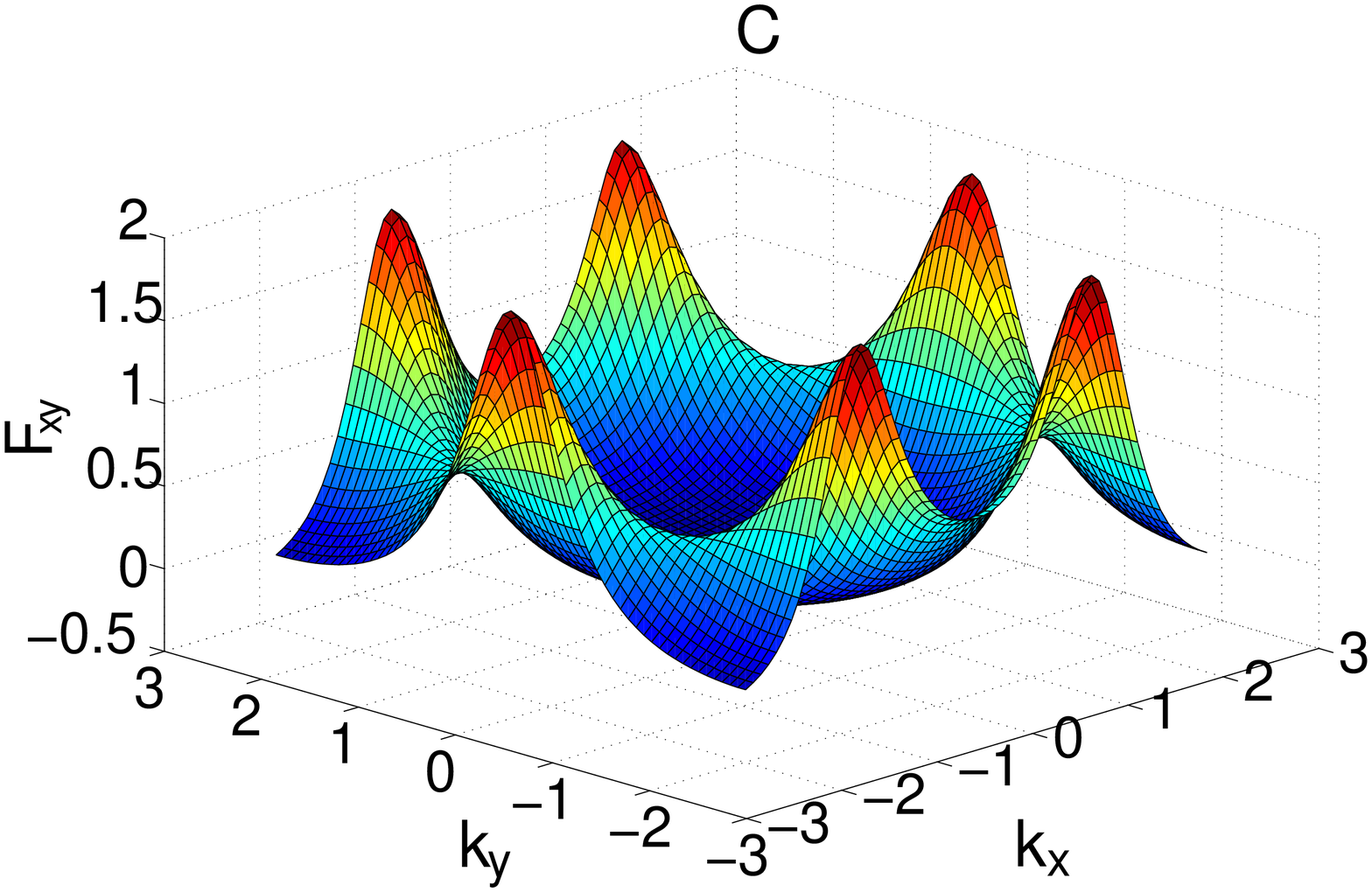,clip=1,width=0.49\linewidth,angle=0}
\centering\epsfig{file=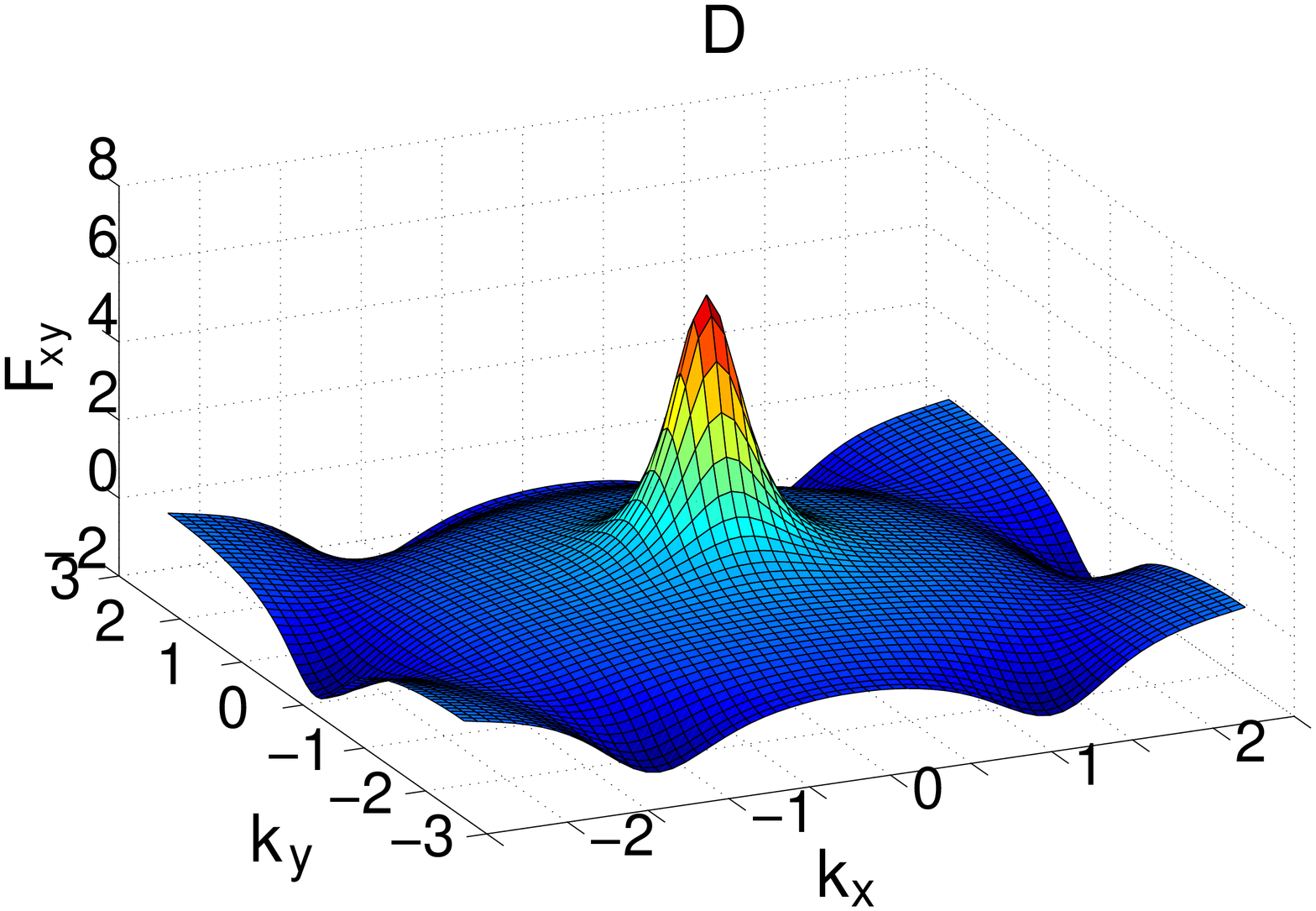,clip=1,width=0.49\linewidth,angle=0}
\centering\epsfig{file=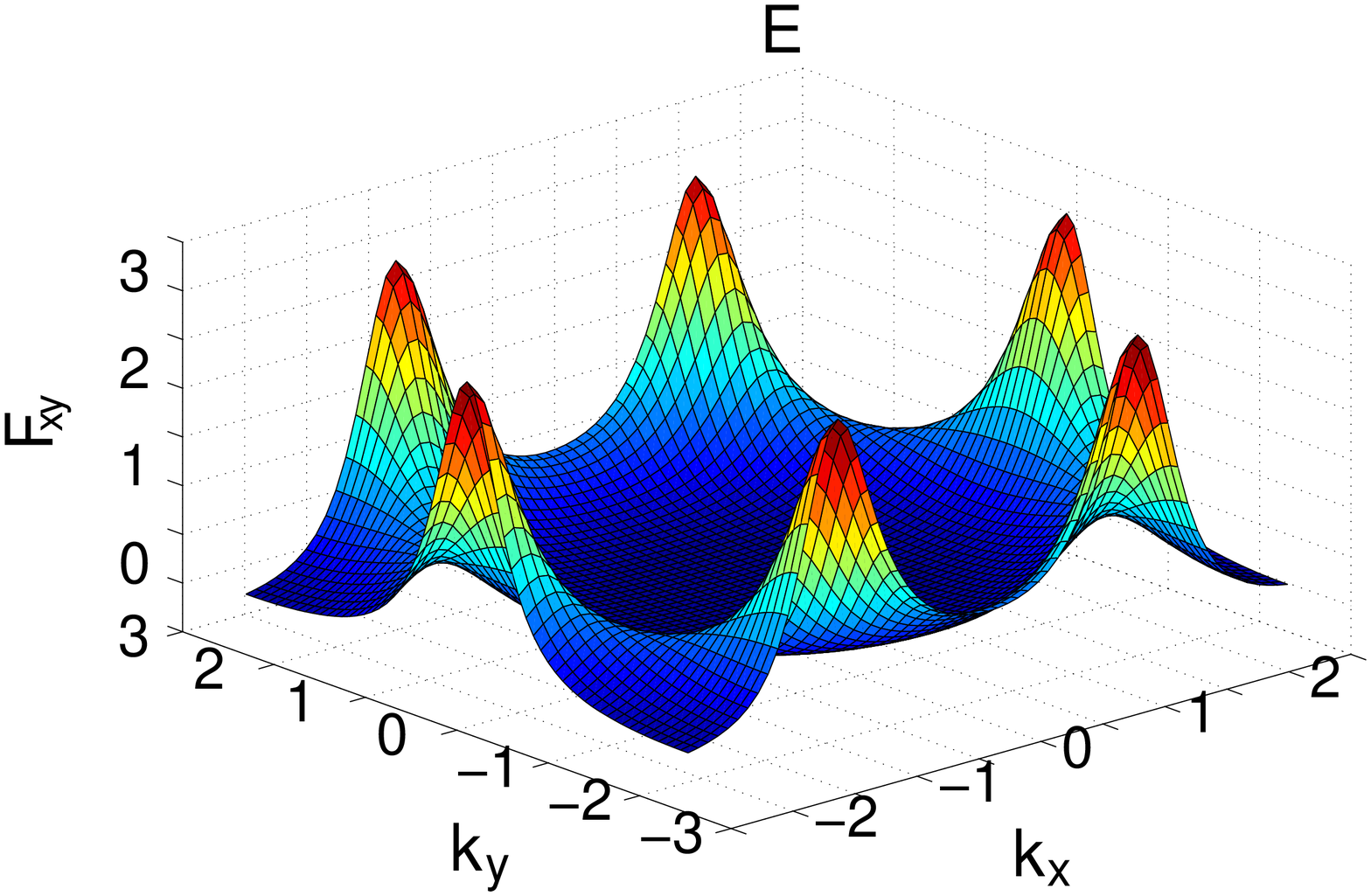,clip=1,width=0.49\linewidth,angle=0}
\centering\epsfig{file=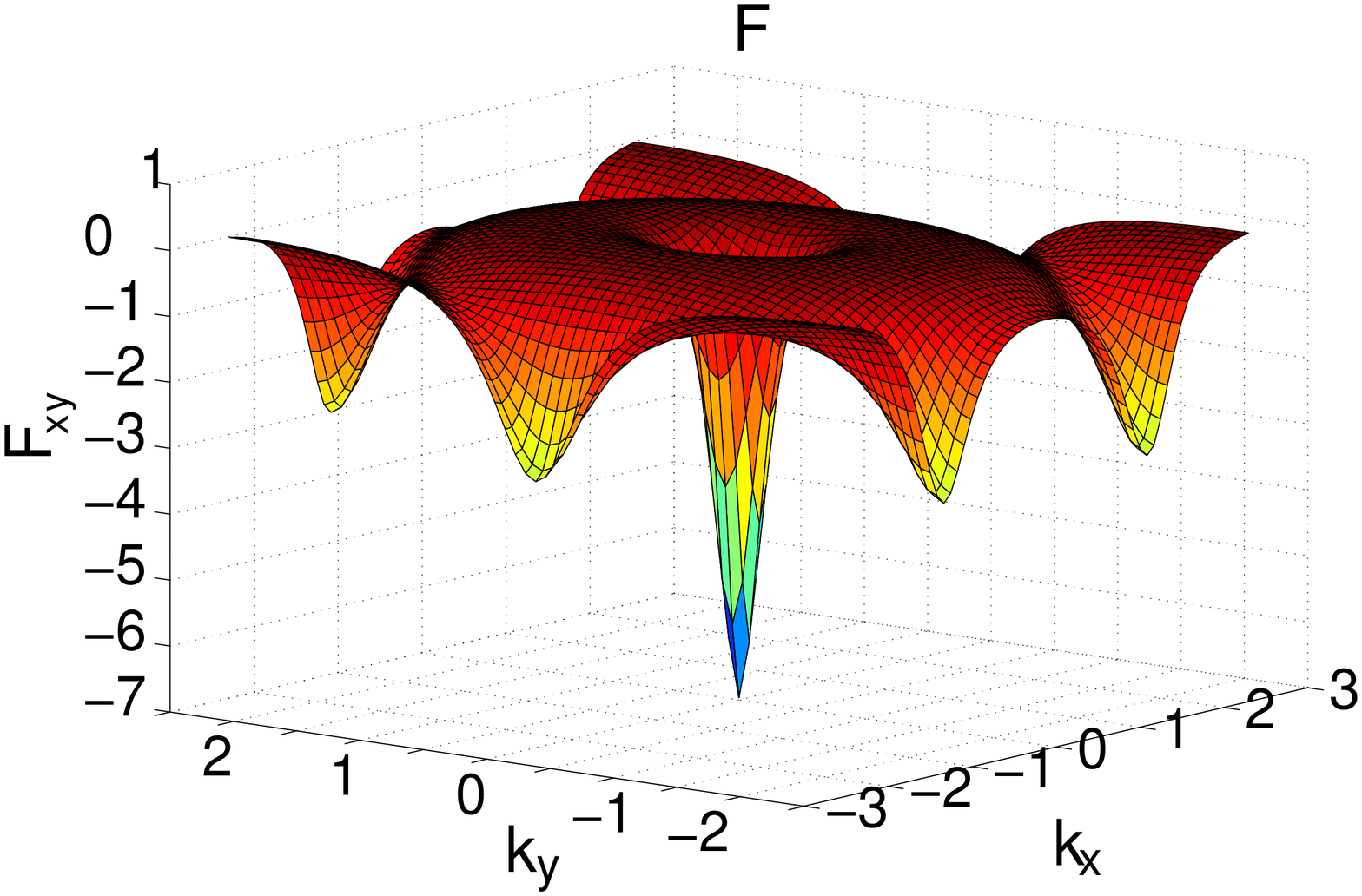,clip=1,width=0.49\linewidth,angle=0}
\caption{The distribution of Berry curvature $F_{xy} (\vec k)$ in the
Brillouin zone for the lower two bands at different $\Omega_z$. 
A, C and E (B, D, and F) are $F_{xy} (\vec k)$ of the first (second)
band at $\Omega_z/t_\pp=0.3$ and $1.7$, respectively.
The Chern number of the first band is 1, and that of the second band
changes from 0 to -1 at $\Omega_z/t_\pp=\frac{3}{2}$.
}
\label{fig:chern}
\end{figure}

The above band structures exhibit non-trivial topological properties.
The Berry curvature $F_{xy}(\vec k)$, or the gauge field strength, 
in the momentum space for the $n$-th band $(n=1\sim 4)$ is defined as
$
F_{n,xy}(\vec k)=\partial_{k_x} A_{n,y} (\vec k) - \partial_{k_y} A_{n,x} 
(\vec k),
$
where  $A_{n,i}(i=x,y)$ is gauge potential defined as
$A_{n,i}=i\avg{\psi_n(\vec k)| \partial_{k_i}| \psi_n (\vec k)}$
\cite{thouless1982,kohmoto1985}.
The eigenstates of the lower two bands are related to those of the
upper two by the transformation
$|\psi_{4-n}(-\vec k)\rangle = (T P)|\psi_n(\vec k)\rangle (n=1,2)$,
thus the Berry curvature satisfies $F_{4-n,xy}(-\vec k)=-F_{n,xy}(\vec k)$.
The field strength $F_{xy}$ of the lower two band is depicted
in Fig. \ref{fig:chern} at different angular velocities.
$F_{n,xy}$ mainly distributes at wavectors $\vec k$ with small gap values 
of $|E_{n\pm 1}(\vec k)-E_{n} (\vec k)|$.
The total flux in the BZ for each band is quantized 
known as the Chern number 
$C_n=\frac{1}{2\pi} \int d^2 k ~ F_{n,xy} (\vec k)$
\cite{thouless1982,kohmoto1985}.
At all values of $\Omega_z>0$, the Chern number of band 1 is
quantized to $1$, in spite of a significant change of distribution 
of $F_{xy}$ as increasing $\Omega_z$ as depicted in Fig. 
\ref{fig:chern} A, C and E.
The maximal of $F_{xy}$ are distributed among a ring around the 
BZ center at small values of $\Omega_z$, and are
pushed to the two vertexes $K_{1,2}$ of BZ as $\Omega_z$
increases.
The Chern number of band 2 is more subtle.
At small $\Omega_z$, each of two massive Dirac points at $K_{1,2}$ 
approximately contribute a flux of $\frac{1}{2}$. 
As $\Omega_z/t_\pp\rightarrow \frac{3}{2} $ from below, the maximum 
of $F_{xy}$ is shifted to the new Dirac point at the BZ center, 
which approximately contributes the flux of $\frac{1}{2}$.
However, these contributions are canceled by the background negative
flux at $\Omega_z/t_\pp<\frac{3}{2}$, and thus the Chern number
is $0$.
A topological quantum phase transition occurs at
$\Omega_z/t_\pp>\frac{3}{2}$ beyond which the flux from the 
Dirac point $K_0$ flips the sign to $-\frac{1}{2}$. 
Combined with the background contribution, the Chern number of
$C_2$ changes to $-1$.
In analogy to electron systems, the transverse conductivity 
can be defined as the ration between the mass flow and
the potential gradient as $\sigma_{xy}= -J_x/\partial_y V$.
% which can be accordingly expressed as
%\bea
%\sigma_{xy}=\frac{m}{\hbar}\frac{1}{2\pi}
%\sum_n \int d^2 k F_{n,xy} (\vec k) n_f(E_n (\vec k)-\mu),
%\eea
%where $n_f$ is the Fermi distribution \cite{nagaosa2006}.
When the Fermi level lies in the band gap, $\sigma_{xy}$ 
is quantized as the sum of
the Chern numbers of the occupied bands.

\begin{figure}
\centering\epsfig{file=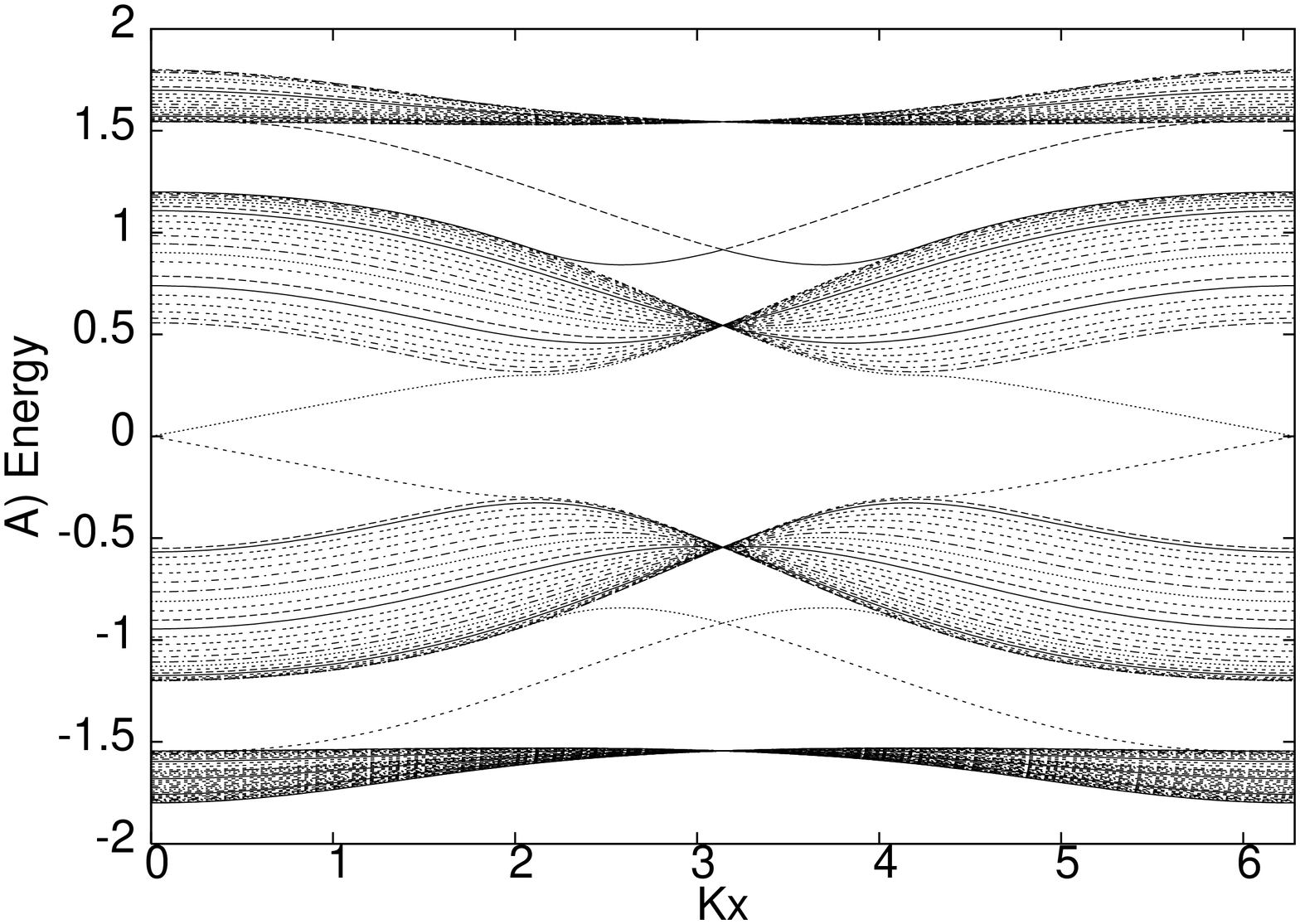,clip=1,width=0.5\linewidth,angle=-90}
\centering\epsfig{file=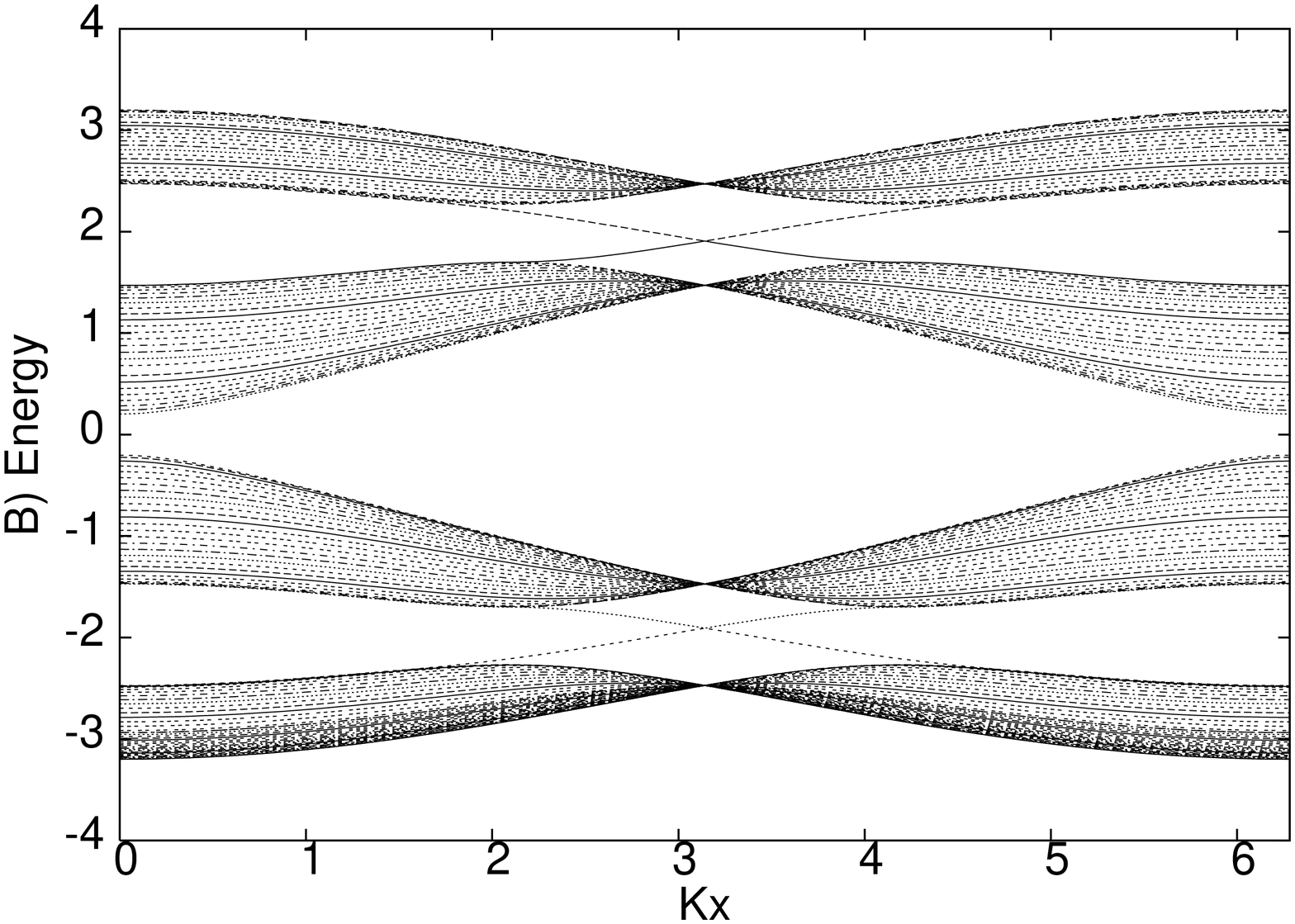,clip=1,width=0.5\linewidth,angle=-90}
\caption{The gapless edge excitations with the open boundary condition 
along the zig-zag edge of the hexagon lattice. 
A) $\Omega_z/t_\pp=0.3$; B) $\Omega_z/t_\pp=1.7$.
A topological phase transition occurs at $\Omega_z/t_\pp=\frac{3}{2}$
above which the edge modes between the middle two bands disappear. 
}
\label{fig:edge}
\end{figure}

The above band structure with non-vanishing Chern numbers gives rise to 
topological stable gapless edge modes lying inside the band gap.
Fig. \ref{fig:edge} depicts the spectra with the open boundary condition
on the zig-zag edges.
The number of chiral edge modes inside the gap between $n$ and $n+1$
band is the sum of Chern numbers from band 1 to $n$, i.e., $\sum_{i=1,n} C_n$.
At $\Omega_z<\frac{3}{2} t_\pp$, the Chern numbers reads 
$C_1=-C_4=-1$ and $C_2=-C_3=0$, thus edge modes exist in all 
of the three band gaps with the same chirality.
At $\Omega_z>\frac{3}{2} t_\pp$, $C_2$ and $C_3$
change to $C_2=-C_3=-1$.
Thus the edge modes between band 1 and 2, and that between band 3
and 4 are of the opposite chiralities.
No edge mode appears between band 2 and 3.
This agrees with the picture that Eq. \ref{eq:ham0} reduces 
to two copies of Haldane's model at $\Omega_z\gg t_\pp$.

\begin{figure}
\centering\epsfig{file=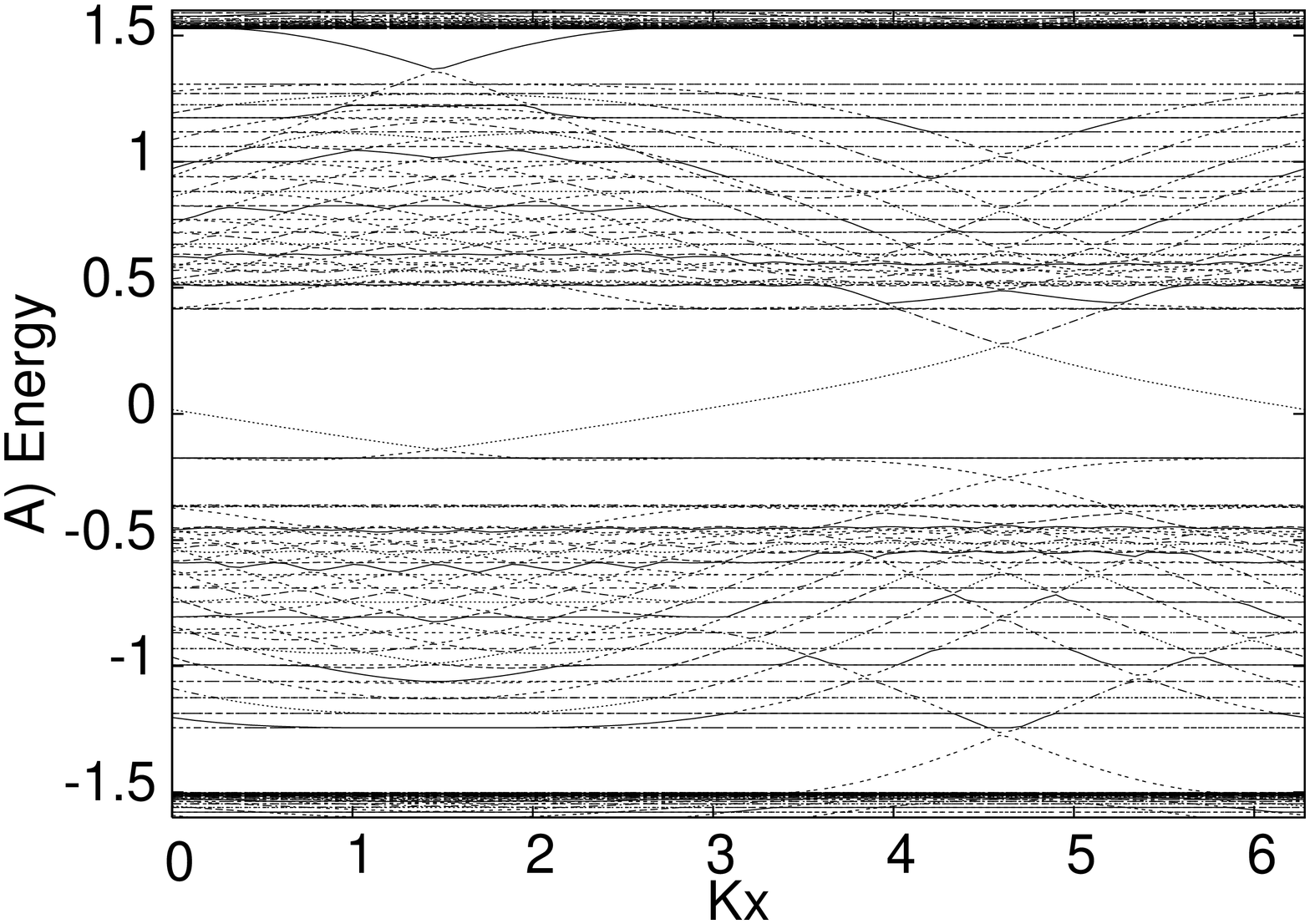,clip=1,width=0.6\linewidth,angle=-90}
\centering\epsfig{file=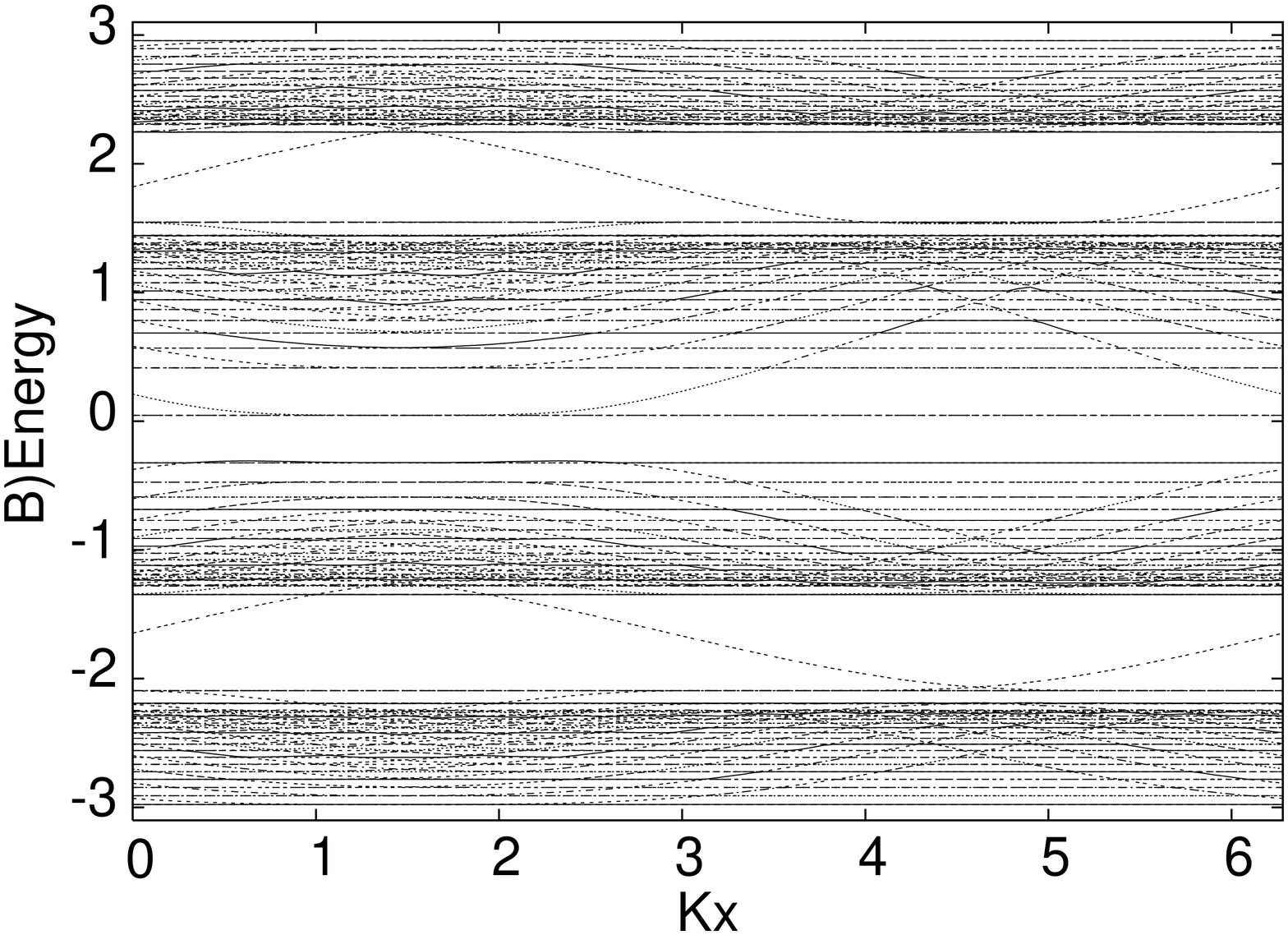,clip=1,width=0.6\linewidth,angle=-90}
\caption{Edge and bulk states spectra of Eq. \ref{eq:QHE} with the open 
boundary condition along the zig-zag edge. The flux per plaquette 
$\phi/(2\pi)=1/20$. A) $\Omega_z/t_\pp=0.2$; B)$\Omega_z/t_\pp=1.5$.
}
\label{fig:QHEedge}
\end{figure}

We also study the QH effect on Eq. \ref{eq:ham0} arising from Landau levels
(LL) by replacing the hopping part to 
\bea
H_{hop}=t_{\pp} \sum_{ \vec{r} \in A}
\big\{ p^\dagger_{\vec{r},i} p_{\vec{r}+a\hat{e}_i,i} 
e^{i\int_{r+ a e_i}^r  \vec A \cdot d \vec r} +h.c. \big\},
\label{eq:QHE}
\eea
where the vector potential-$\vec A$ can be generated through another 
overall lattice rotation or by light induced gauge potential.
We will take the flux per plaquette $\Phi$ and $\Omega_z$
as two independent variables.
The spectra of the above Hamiltonian does not depend on the gauge choice
but the physical wavefunctions differ by a gauge transformation.
For the calculation convenience, we use the Landau gauge for an open 
boundary system along the zig-zag edge and take $\Phi/(2\pi)=0.05$.

Due to the vector potential $\vec A$, $(TP) (H_{hop}+H_L) 
(TP)^{-1} \neq -(H_{hop}+H_L) $, thus the spectra are no longer symmetric
respect to the zero energy.
Generally speaking, all of the four bands split into a number of
flat LLs with dispersive edge modes lie in between.
The pattern of edge modes does not change much as varying the value 
of $\Phi$, but significantly changes as increasing $\Omega_z$.
At small values of $\Omega_z$ (e.g., $\Omega_z =0.2 t_\pp$ as
shown in Fig. \ref{fig:QHEedge}. A ), gapless edge modes go through 
the entire spectra from the very band bottom to top.
Landau levels close to the zero energy arise from Dirac cones at $K_{1,2}$
with opposite masses as shown in Eq. \ref{eq:dirac1}.
The $0$th LL is pushed to negative energy at the gap value
around $-0.26t_\pp$.
The number of chiral edge modes between levels of $n=0$ to $\pm1$
is 1 with opposite chirality and that between $n=\pm 1$ and $\pm 2$ is 3.
The energies of $n=\pm1$ and $n=\pm 2$ appear roughly symmetric to 
zero energy. 
Next let us look at $\Omega_z=\frac{3}{2} t_\pp$ where a single gapless
Dirac cone appears as shown in Eq. \ref{eq:dirac2}.
Indeed the $0$th LL appear close to the zero energy but with a small
deviation, which is understandable as no exact symmetry to protect it
right at the zero energy.
It is tempting to think the appearance of the half-integer QH effect, 
but this is impossible in free lattice fermion systems \cite{haldane1988}.
Another half has to be contributed from the high energy part of the
band structure.
As a result, the number of chiral edge modes between LLs $n=0$
and $1$ is $\frac{1}{2}+\frac{1}{2}=1$, while that between LLs 
$n=0$ and $-1$ is $-\frac{1}{2}+\frac{1}{2}=0$.
Thus the spectra from bottom to top become disconnected without
edge modes connecting them.
This disconnection actually begins to appear even earlier at
$\Omega_z=1.2 t_\pp$, and is enhanced as $\Omega_z$ goes larger.
At large values of $\Omega_z$, the model reduces to two copies
($p_x\pm ip_y$) of Haldane's model.
The patterns of LLs between band 1 and 2, and between band 3 and 4
become the those of the two massive Dirac cones with opposite
mass signs.
When Fermi level lies in between LLs, the transverse conductance
$\sigma_{xy}$ is quantized at the number of chiral
edge modes.

In summary, we propose to investigate the topological insulating 
states in the $p$-orbital systems in the honeycomb lattice, which 
can be realized by the current available experimental techniques.
The orbital angular momentum polarization generates non-trivial
Chern numbers in the band structure, which gives rise to
the orbital counterpart of QAH effect without LLs.
QH effect arising for LLs are also investigated, which shows
quantitative different features from those in graphene.

C. W. thanks D. Arovas, M. Fogler and J. Hirsch for helpful discussions,
and N. Gemekel for the introduction the method of rotating optical lattices.
C. W. is supported by the start-up funding at UCSD, and the
Sloan Research Foundation.

%\bibliographystyle{prsty}
%\bibliography{orbital,spin32,extra,topo}

\end{document}